\documentclass[runningheads]{svmult}
\usepackage{makeidx}   
\usepackage{graphicx}  
\usepackage{subeqnar}  
\usepackage{multicol}  
\usepackage{cropmark} 
\usepackage{physprbb}  
\newcommand{\hdmo}{Heidelberg-Moscow experiment}

\newcommand{\ch}[2]{$\rm ^{#1}#2 $}
\newcommand{\bea}{\begin{eqnarray}}
\newcommand{\eea}{\end{eqnarray}}

\newcommand{\lro}{\left(}
\newcommand{\rro}{\right)}

\def \beq {\begin{equation}}
\def \eeq {\end{equation}}

\def \gs  {Gran Sasso underground laboratory} 
\def \onbb {$0\nu\beta\beta$ }
\def \tnbb {$2\nu\beta\beta$ }
\def \mnbb {$0\nu\chi\beta\beta$ }

\begin{document}
\title*{Latest Results from the Heidelberg-Moscow Double-Beta-Decay 
Experiment\thanks{Talk presented by A. Dietz at the Third International 
Conference {\it DARK2000},\\ Heidelberg, GERMANY, July 10-15, 2000}}
\toctitle{Latest results from the {\bf Heidelberg-Moscow} Double-Beta-Decay 
experiment}
%
%
\titlerunning{Latest results from the {\bf Heidelberg-Moscow} experiment}
%
\author{H.V. Klapdor-Kleingrothaus\inst{1} \inst{,3}
\and A. Dietz\inst{1}
\and L. Baudis\inst{1}
\and G. Heusser\inst{1} 
\and\\ I.V. Krivosheina\inst{1}
\and S. Kolb\inst{1}
\and B. Majorovits\inst{1}
\and H. Paes\inst{1}
\and H. Strecker\inst{1}
\and\\ V. Alexeev\inst{2}
\and A. Balysh\inst{2}
\and A. Bakalyarov\inst{2}
\and S.T. Belyaev\inst{2}
\and V.I. Lebedev\inst{2}
\and\\ S. Zhukov\inst{2}}
\authorrunning{H.V. Klapdor-Kleingrothaus {\sl et. al.}}
%
%
\institute{Max-Planck-Institute f\"ur Kernphysik\\
 Postfach 10 39 80, D-69029 Heidelberg\\
\and Russian Science Centre, Kurchatov Institute,\\
     123 182 Moscow, Russia\\
\and Spokesman of HEIDELBERG-MOSCOW and GENIUS Collaborations.\\
E-mail:klapdor@gustav.mpi-hd.mpg.de. Home Page Heidelberg Non-Accelerator 
Particle Physics group: http://mpi-hd.mpg.de.non$\_$acc/}

\maketitle              

\begin{abstract}
New results for the double beta decay of \ch{76}{Ge} are presented. 
They are extracted from Data obtained with the {\hdmo}, which operates
five enriched \ch{76}{Ge} detectors in an extreme low-level
environment in the {\gs}. 
The two neutrino accompanied double beta decay is evaluated for the
first time for all five detectors with a statistical significance of
47.7\,kg\,y resulting in a half life of
T$_{1/2}^{2\nu}$\,=\,[1.55$\pm$0.01(stat)\,$^{+0.19}_{-0.15}$(syst)]$\times$ 
10$^{21}$\,years. 
The lower limit on the half-life of the {\onbb}-decay obtained with
pulse shape analysis is  \mbox{T$_{1/2}^{0\nu} >$
1.9$\times$10$^{25}$ (3.1$\times$10$^{25}$)} years with 90\%
C.L. (68\% C.L.) (with 35.5\,kg\,y).
This results in an upper limit of the effective Majorana neutrino mass 
of 0.35\,eV (0.27\,eV).
No evidence for a Majoron emitting decay mode or for the neutrinoless
mode is observed. 
\end{abstract}

\section{Double Beta Decay}
There seems to be a general consensus over the neutrino oscillation
interpretation of the atmospheric and solar neutrino data,  
delivering a strong indication for a non-vanishing neutrino mass.
While such kind of experiments yields information on the difference of 
squared neutrino mass eigenvalues and on mixing angles, the 
absolute scale of the neutrino mass is still unknown.
Information from double beta decay experiments is indispensible to
solve these questions \cite{paes,kk1,kk2,kk3}.
Another important problem is that of the fundamental
character of the neutrino, whether it is a Dirac or a Majorana
particle.
Neutrinoless double beta decay could answer also this question. The
HEIDELBERG-MOSCOW experiment is giving since almost eight years now, the
most sensitive limit of all $\beta\beta$-experiments worldwide \cite{kk3}.
Double beta decay, the rarest known nuclear decay process, can occur
in different modes:

\begin{tabbing}
\hspace*{4.5cm} \= \hspace*{7.9cm} \= \kill
\qquad $ 2\nu\beta\beta- \rm decay:$ \> $ A(Z,N) \rightarrow
A(Z\!+\!2,N\!-\!2) + 2e^- + 2\bar{\nu}_e$ \hspace*{1.05cm}(1)\\[-0ex]
\qquad $ 0\nu\beta\beta- \rm decay:$ \> $ A(Z,N) \rightarrow
A(Z\!+\!2,N\!-\!2) + 2e^-$ \hspace*{2cm}(2) \\[-0ex]
\qquad $ 0\nu(2)\chi\beta\beta- \rm decay:$ \> $ A(Z,N) \rightarrow
A(Z\!+\!2,N\!-\!2) + 2e^- + (2)\chi$ \hspace*{0.9cm}(3) \\[-0ex]
\end{tabbing}

The two-neutrino decay mode (1) is a conventional second order weak 
process, allowed in the Standard Model of particle physics. 
So far it has been observed for about 10 different
nuclei \cite{kk2,kk3,vogel}. 
An accurate  measurement of the half-life of the decay is of
importance, since it provides a cross-check on  the reliability of 
matrix element calculations.
The Majoron emitting decay mode (2) could reveal the existence of 
light or massless bosons, so called Majorons, with a non-zero
coupling to neutrinos.
The neutrinoless mode (3) is by far the most exciting one due to the
violation of the lepton number by two units. It can not only probe a
Majorana neutrino mass, but various new physics scenarios beyond the
Standard Model, such as R-parity violating supersymmetric models
\cite{rpviol}, R-parity conserving SUSY models \cite{rpcons}, leptoquarks
\cite{leptoquarks}, violation of Lorentz-invariance \cite{lorinvariance} and
compositeness \cite{compo} (for a review see \cite{kk3,kk98,kklb,kk00}). Any theory containing lepton number violating
interactions can in principle lead to this process allowing to obtain
information on the specific underlying theory.
The experimental signature of the neutrinoless mode is a peak at the
Q-value of the decay, whereas for the two-neutrino and
Majoron-accompanied decay 
modes well defined continuous energy spectra are expected.
They are identified by their spectral index n, defined as the power of
the energy in the phase space integral (see \cite{bamert}).
The Majoron emitting modes are characterized by n=1,3,7, while 
for the 2$\nu\beta\beta$ decay, n=5.

\section{The \hdmo}

The \hdmo{} operates five p-type HPGe detectors in the Gran Sasso
underground laboratory which were originally grown from 19.2\,kg of enriched \ch{76}{Ge}. 
The total active mass of the detectors is 10.96\,kg, corresponding to
125.5\,mol of \ch{76}{Ge}, the presently largest source strength of
all double beta experiments. 
The enrichment of the used Germanium is 86\%. 
A detailed description of the experiment is given in \cite{hdmo}.

To check the stability of the experiment, a couple of parameters such
as temperature, nitrogen flow, leakage current of the detectors,
overall and individual trigger rates are monitored daily. An energy
calibration is done weekly with a \ch{228}{Th} and a \ch{152}{Eu}-\ch{228}{Th}
source. The energy resolution of the detectors at 2614\,keV ranges from
 3--3.7\,keV.
The energy thresholds for data recording are set to about 70\,keV
(with exception of the second detector, which is used for 
dark matter measurements in addition, see \cite{prd}).

\begin{figure}[h]
\begin{center}
\includegraphics[height=5cm]{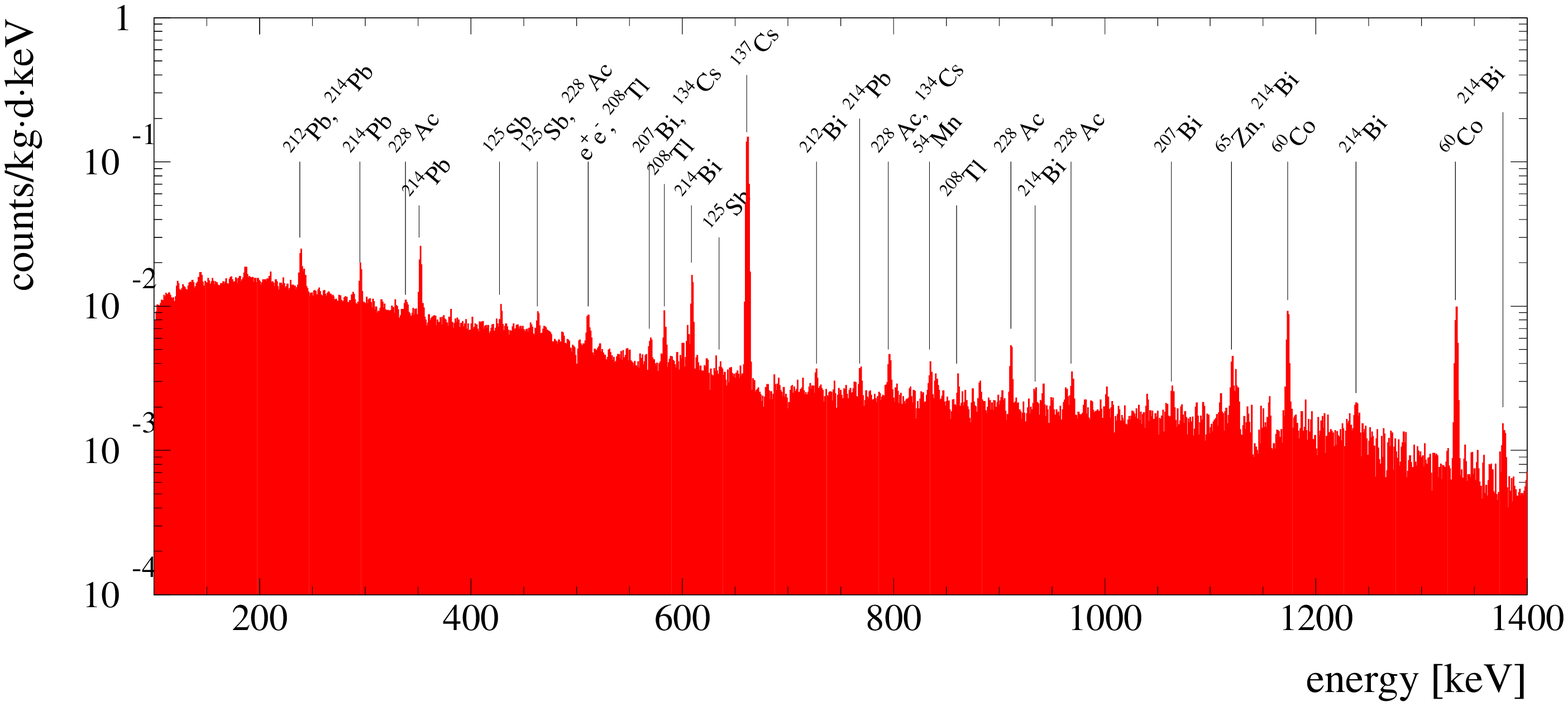}
\includegraphics[height=5cm]{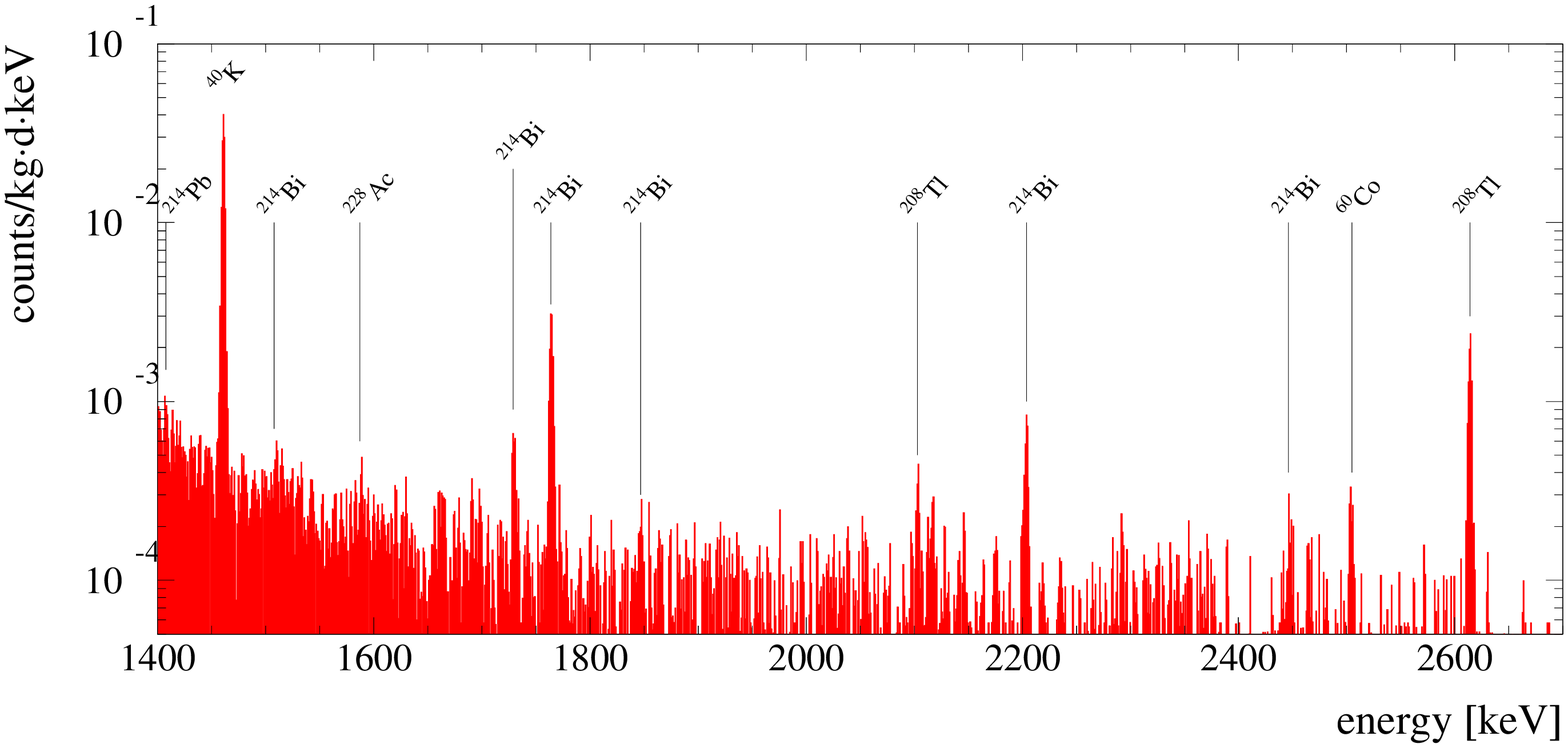}
\end{center}
\caption{\label{sumspectrum}Sum spectrum of all five $^{76}$Ge
  detectors after 47.4~kg~y of measurement. The most prominent
  identified lines are labeled.}
\end{figure}

Figure~\ref{sumspectrum} shows the combined sum spectrum of all five
enriched detectors of the \hdmo{} with a statistical significance of 
47.4\,kg\,y (see \cite{dietz}). 
The large peak-to-Compton ratio of the detectors facilitates the
identification of $\gamma$ activities. 
The easily identified background components
consist of primordial activities of the natural
decay chains from \ch{238}{U} and \ch{232}{Th} from \ch{40}{K}, 
anthropogenic radio nuclides, like \ch{137}{Cs}, \ch{134}{Cs}, 
\ch{125}{Sb} and \ch{207}{Bi} and cosmogenic isotopes, such as 
 \ch{54}{Mn}, \ch{57}{Co}, \ch{58}{Co},  \ch{60}{Co} and \ch{65}{Zn}.
Hidden in the continuous background are the contributions of  the
brems\-strahlungs spectrum of \ch{210}{Bi} (daughter of \ch{210}{Pb}),
elastic and inelastic neutron scattering and direct muon
induced events.  \\

\section{Background model}

\begin{figure}[t]
\hspace*{-1.5cm}
\includegraphics[height=10cm]{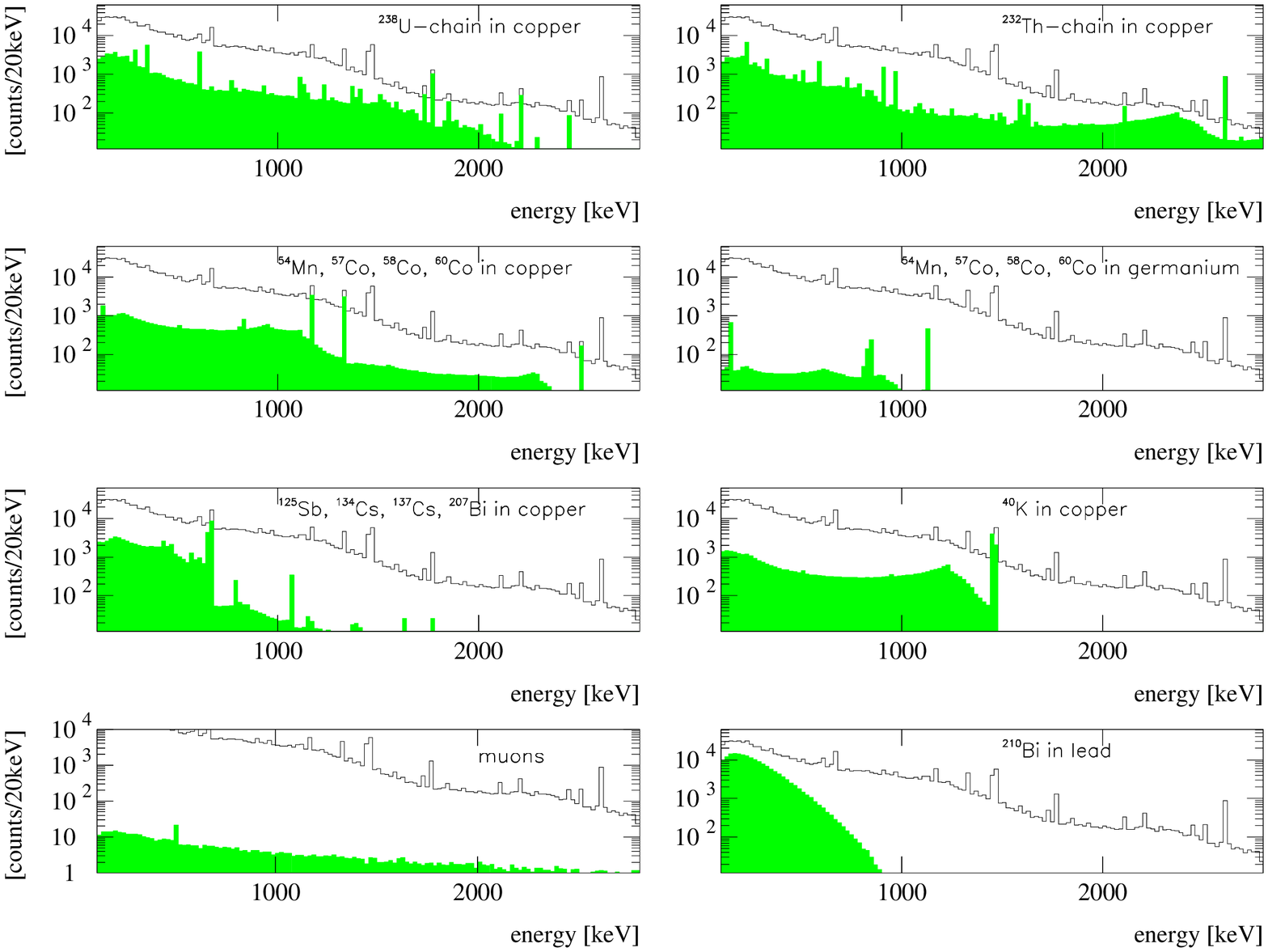}
\caption{\label{simspec}The simulated background components (shaded
  areas) compared with the original measured sum spectrum for all five detectors.}
\end{figure}

The evaluation of the spectra caused by the \tnbb{} decay and the
Majoron-emitting decay modes requires a detailed knowledge of the
composition of the background on which they are superimposed.
To unfold the background, a Monte Carlo simulation was performed. It is
based on the CERN code {\sc Geant3.21}, modified for simulating
radioactive decays with the complete implemented decay schemes taken from
\cite{decayscheme}.
Five parts of the experimental setup have been identified to represent 
the main locations of the 
radioactive impurities: the LC2-Pb shield, the copper shield, the
copper and plastic parts of the cryostats and the Ge crystals themselves. 
Other materials or locations in the detector array are negligible due their small masses and low activities.
The following background components were simulated: 
the natural decay chains of \ch{238}{U} and \ch{232}{Th}, \ch{40}{K},
cosmogenic and anthropogenic isotopes, muon showers and neutron
induced interactions.
It was assumed that the \ch{238}{U} and \ch{232}{Th} decay chains are
in secular equilibrium and that the radioactive isotopes in the
respective materials are uniformly distributed.
Muon-induced showers were simulated based on the measured flux and
energy distribution of muons in the \gs{} \cite{macro}.
Not considered were muon-induced neutrons in the detector shielding
materials, due to still large uncertainties in the absolute n-flux
determinations in {\sc Geant3.21}.
This component belongs to the non-identified background 
which will be discussed below.
The measured neutron flux in the \gs{} \cite{belli} was simulated using the
MICAP implementation in {\sc Geant} \cite{micap}.
The activities of \ch{40}{K} and \ch{210}{Pb} in the LC2-Pb shield
were determined in  separate activity measurements \cite{pernicka}.
In order to extract the best fit values for each activity,
a least-squares method has been used.
The location of the radioactive impurities was determined by comparing
the peak intensities of multiline isotopes with the simulation.
The error of a possible misplacement is part of the systematic error
of the background model. 
The influence of each radioactive impurity located in one
detector on all other detectors was considered.
We identified a total number of 142 lines in the spectra of the 
five enriched Ge detectors. Their measured intensities were used to 
normalize the simulated components of the background model.
Table~\ref{activities} shows the identified background components,
their estimated activities and their most probable locations in the
experimental setup.
The main background sources (natural decay chains, cosmogenics and
anthropogenic radionuclides) were located in the copper parts of the
cryostats. In the Ge crystals themselves, only cosmogenic radionuclides
were identified. 
There is no intrinsic U/Th contamination of the
crystals, due to the absence of $\alpha$-peaks in their high energy
spectra (the single $\alpha$-line at 5.3\,MeV detected in two of the
five detectors originates most likely from surface contaminations at
the inner contact). 
External $\alpha$ and $\beta$ activities are shielded by the about 0.7
mm inactive zone of the p-type detectors on the outer crystal surface.
Figure \ref{simspec} shows the contribution of the simulated
background  components on the original measured sum spectrum of the Ge
detectors (for details of the simulations see \cite{dietz}). 

\begin{table}[t]
\begin{center}
\renewcommand{\arraystretch}{1.4}
\setlength\tabcolsep{3pt}
\begin{tabular}{l|l|c|l|l|c}
Isotope & \multicolumn{2}{c|}{Average for all 5 detectors}&
Isotope & \multicolumn{2}{c}{Average for all 5 detectors} \\ 
        & Localisation & Activity& & Localisation & Activity \\
        &              & [$\mu Bq/kg]$ &              && [$\mu Bq/kg]$\\ \hline
\ch{238}{U}  & Cu cryostat  & 85.0              & \ch{65}{Zn} & Ge crystal & 20.2 (no. 2-4)\\
\ch{238}{U}  & Pb shield    & $<$11.3           &  \ch{54}{Mn} & Cu cryostat   & 17.1\\
\ch{232}{Th} & Cu cryostat  & 62.5              & \ch{57}{Co} & Cu cryostat   & 32.4\\
\ch{232}{Th} & Pb shield    & $<$0.9            & \ch{58}{Co} & Cu cryostat   & 23.4 (only no. 3-5)\\
\ch{40}{K}  & Cu cryostat   & 480.3             &  \ch{60}{Co} & Cu cryostat   & 65.2\\
\ch{40}{K}  & LC2-Pb        & 310 (ext. meas.)  & \ch{125}{Sb}& Cu cryostat   & 36.2\\
\ch{210}{Pb} & LC2-Pb   & 3.6$\times$10$^5$ (ext. meas.)  &\ch{134}{Cs} &Cu cryostat   & 5.1 \\ 
\ch{54}{Mn} & Ge crystal & 4.2                  & \ch{137}{Cs}    & Cu cryostat   & 67.8 (no.5: 463.9)\\
\ch{57}{Co} & Ge crystal & 2.6                  &  \ch{207}{Bi}& Cu cryostat   & 7.2 \\
\ch{58}{Co} & Ge crystal & 3.4 (only no. 3 \& 5)&&\\
\end{tabular}
\end{center}
\caption{\label{activities}Identified background components
  (primordial, cosmogenic, anthropogenic), their estimated activities
  and most probable locations in the full setup of the
  Heidelberg-Moscow experiment.} 
\end{table}

\section{Results for the 2$\nu\beta\beta$ and the 0$\nu(\chi)\chi\beta\beta$ 
  decays}
\begin{figure}[t]
\begin{center}
\includegraphics[height=8cm]{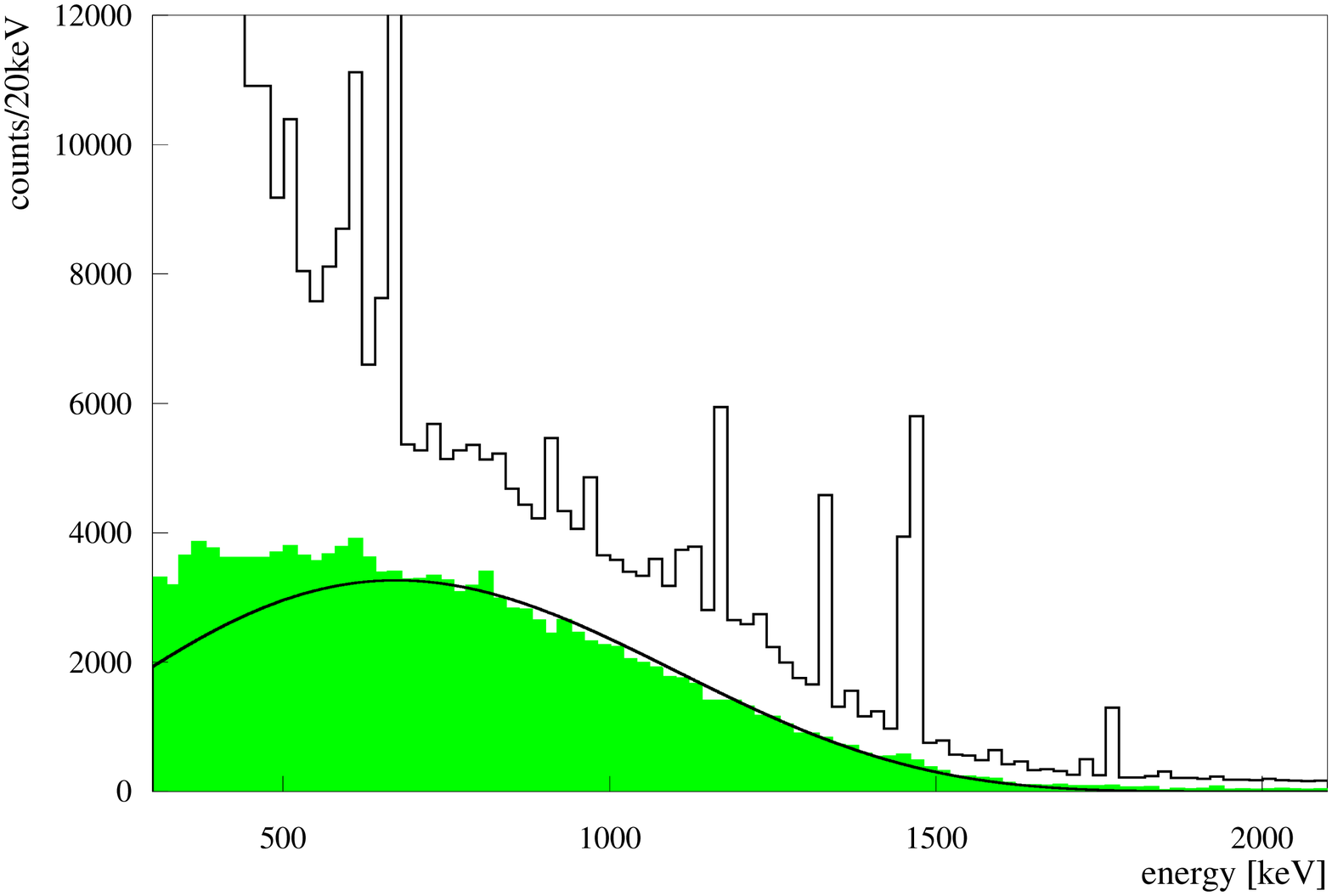}
\end{center}
\caption{\label{total}Summed spectra of all five detectors after 47.7
  kg y of measurement together
  with the residual spectrum after subtracting all identified background
  components. The thick line shows the fitted  2$\nu\beta\beta$-signal.}
\end{figure}

In Fig. \ref{total} the summed data of the five detectors are shown
together with the result after subtracting the identified background
components. A bin width of 20\,keV  is chosen in order to avoid 
statistical fluctuations when subtracting the simulated $\gamma$
lines from the measured spectrum.
The contribution of the {\tnbb} decay to the residual spectrum is
clearly visible. Its half-life was determined under 
the assumption that the entire residual
spectrum is composed of the {\tnbb}-signal. 
Due to non-identified background in the energy region below 700\,keV,
the fit interval for the  $2\nu\beta\beta$-signal is chosen between 
700--2040\,keV. With the above assumption, this region contains 64553 
\tnbb{} events, corresponding to 51.7\% of the total \tnbb{}-signal.

The theoretically expected \tnbb{} spectrum was fitted to the data in a
maximum-likelihood-fit with T$_{1/2}$ as free parameter, resulting in
the following half-life for the \tnbb{}-decay at  68\% C.L. (combined
result for the five detectors):

\setcounter{equation}{3}
\beq
T^{2\nu}_{\frac{1}{2}}= \lro \; 1.55\pm0.01 (\rm stat)\;^{+0.19}_{-0.15}(\rm syst) \;\rro \times 10^{21}\, {\rm y}
\eeq

The statistical error is evaluated from the parabolic behaviour of the
logarithmic likelihood ratio which corresponds to a $\chi^2$
function. The systematic error includes the error of the simulated
 detector response, the error made by the misplacement of background activities
 and the normalization error due to the statistical error of the measured   
  $\gamma$-lines (see \cite{dietz}).

The inferred value for the half-life is consistent with earlier
results of this experiment \cite{hdmo,balysh} and with the result of 
\cite{igex}, as well as with the range of theoretical predictions,
which lie between 1.5$\times 10^{20}$\,--\,2.99 $\times 10^{21}$\,y
\cite{muto,caurier,engel,wu,staudt}. The prediction of
\cite{muto,staudt} for the 2$\nu\beta\beta$ matrix element agrees
within a 
factor of $\sqrt{2}$ with the experimental value.

\begin{table}[t]
\begin{center}
\renewcommand{\arraystretch}{1.4}
\setlength\tabcolsep{8pt}
\begin{tabular}{l|l|l|l|l}
Modus & Model & n & T$_{1/2}^{0\nu\chi} > $ (90\% C.L.) & $\langle
g_{\nu\chi} \rangle \,< $ (90\% C.L.)\\
\hline
$\chi\beta\beta$ & \cite{chi80} & 1 & 6.4$\times$10$^{22}$\,y & 8.1$\times$10$^{-5}$ \\
l$\chi\beta\beta$ & \cite{bur93},\cite{car93} & 3 &
1.4$\times$10$^{22}$\,y & 0.11 (0.04) \\
\end{tabular}
\end{center}
\caption{\label{tabmajoron}Half-life limits for the Majoron-emitting
  decay-modes and derived coupling constants
  using the matrix elements from \cite{hirsch} for
  different majoron models (n is the spectral index of the decay mode). }
\end{table}

The half-life limits of the Majoron-emitting decay-modes were
determined from the same data set by fitting  the \tnbb{} and the
\mnbb{} spectra simultaneously. The considered Majoron models are
described  in \cite{hirsch}. Since the selected  energy interval
starts at 700\,keV, an analysis of the decay-mode with the spectral
index n=7 (maximum at about 500\,keV) was not possible. The results of
the fits for n=1 and n=3 are shown in Table \ref{tabmajoron}.
The \tnbb{} half-lifes extracted in the two-parameter fits are
consistent within 1$\sigma$ with the exclusive double beta decay
evaluation. In Table \ref{maj_comp}  a comparison of the effective Majoron
neutrino couplings extracted for different double beta nuclei is made.  

\begin{table}[t]

\begin{center}
\renewcommand{\arraystretch}{1.4}
\setlength\tabcolsep{8pt}
\begin{tabular}{l|l|l|l|l}
Nucleus & Ref. & T$_{1/2}^{0\nu\chi} > $  & $\langle
g_{\nu\chi} \rangle \,< $ & C.L. [\%]\\
\hline
$^{76}$Ge & this work & 6.4$\times$10$^{22}$\,y & 8.1$\times$10$^{-5}$& 
90\\
$^{82}$Se & \cite{arnold98} & 2.4$\times$10$^{21}$\,y &
2.3$\times$10$^{-4}$ & 90\\
$^{96}$Zr & \cite{arnold99} & 3.5$\times$10$^{20}$\,y &
2.6$\times$10$^{-4}$& 90 \\
$^{100}$Mo & \cite{ejiri96} & 5.4$\times$10$^{21}$\,y & 7.3$\times$10$^{-5}$& 
68\\
$^{116}$Cd & \cite{danevich00} & 3.7$\times$10$^{21}$\,y & 1.2$\times$10$^{-4}$&90 \\
$^{128}$Te & \cite{berna92} & 7.7$\times$10$^{24}$\,y &
3.0$\times$10$^{-5}$ & 90\\
$^{136}$Xe & \cite{luescher98} & 7.2$\times$10$^{21}$\,y &
2.0$\times$10$^{-4}$& 90 \\
$^{150}$Nd & \cite{desilva97} & 2.8$\times$10$^{20}$\,y & 9.9$\times$10$^{-5}$
& 90\\
\end{tabular}
\end{center}
\caption{\label{maj_comp}
Half-life limits on the Majoron-emitting decay-mode {\mnbb} extracted from
different nuclei and the derived limits on the
effective Majoron-neutrino coupling for n=1.}
\end{table}

\section{Results for the 0$\nu\beta\beta$ decay}

For the evaluation of the \onbb{} decay we consider 
the raw data of all five detectors as well as data with pulse shape analysis.
The pulse shape analysis method used here is described elsewhere \cite{hdmo97}.
No further data manipulation is done, e.g. the previously established
background model is not subtracted. We see in none of the two data
sets an indication for a peak at the Q-value of 2038.56$\pm$0.32\,keV
\cite{hykawy} of the \onbb{} decay. 

The total spectrum of the five detectors with a statistical
significance of 53.9\,kg\,y contains all the data with the exception of
the first 200\,d of measurement of each detector, because of possible
interference with the cosmogenic $^{56}$Co. The interpolated energy
resolution at the energy at the hypothetical $0\nu\beta\beta$ peak is
(4.23$\pm$0.14) keV.
The expected background in the \onbb{} region is estimated from the
energy interval 2000--2080\,keV. In this range the background is
(0.19$\pm$0.01)~counts/(kg\,y\,keV). 
The expected background in the 3$\sigma$ peak interval, centered at
2038.56\,keV interpolated from the adjacent energy regions, is
(110.3$\pm$3.9) events. The number of measured events in the same peak 
region is 112. 
To extract a half-life limit for the \onbb{}-decay we follow the
conservative procedure recommended in \cite{pdg96}.

With the achieved energy resolution, the number of excluded events in
the 3$\sigma$ peak region is 19.8 (12) with 90\% C.L. (68\% C.L.),
resulting in a half-life limit of (for the 0$^+ \rightarrow$ 0$^+$
transition):
\begin{eqnarray*}
{\rm T}_{1/2}^{0\nu} \geq 1.3 \times 10^{25} {\rm~ y} \;\;\; 90\% {\rm~ C.L.}\\
{\rm T}_{1/2}^{0\nu} \geq 2.2 \times 10^{25} {\rm~ y} \;\;\; 68\% {\rm~ C.L.}
\end{eqnarray*}

We consider now the data for which the pulse shape of each interaction 
of the detectors was recorded and analyzed. 
The total statistical significance is 35.5\,kg\,y and 
the background index in the energy region between 2000--2080\,keV is
(0.06$\pm$0.01) events/(kg\,y\,keV), about a factor 3 lower than for
the full data set. This  is due to the large fraction of multiple Compton 
scattered events in this energy region, which are effectively
suppressed by the pulse shape discrimination method.
The expected number of events from the background left and right of
the peak region is (20.4$\pm$1.6) events, the measured number of
events in the 3$\sigma$ peak region is 21.
Following again the method proposed by \cite{pdg96}, we can exclude 9.3 (5.5) events with 90\%~C.L. (68~\% C.L.). The limit on the half-life is:
\begin{eqnarray*}
{\rm T}_{1/2}^{0\nu} \geq 1.9 \times 10^{25} {\rm~ y} \;\;\; 90\% {\rm~ C.L.}\\
{\rm T}_{1/2}^{0\nu} \geq 3.1 \times 10^{25} {\rm~ y} \;\;\; 68\% {\rm~ C.L.}
\end{eqnarray*}

\begin{figure}[t]
\begin{center}
\includegraphics[height=8.5cm]{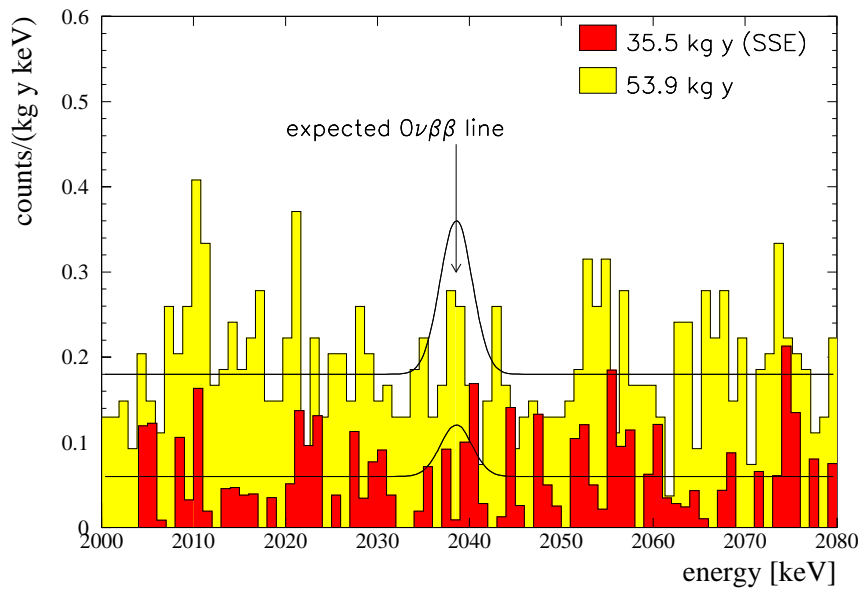}
\end{center}
\caption{Sum spectrum of all five detectors with 53.9\,kg\,y and SSE
  spectrum with 35.5\,kg\,y in the region of interest for the
  \onbb{}-decay. The curves correspond to the excluded signals with
  ${\rm T}_{1/2}^{0\nu} \geq 1.3 \times 10^{25} {\rm~ y}$ (90\%
  C.L.) and ${\rm T}_{1/2}^{0\nu} \geq 1.9 \times 10^{25} {\rm~ y}$
  (90\% C.L.), respectively. }
\label{both}
\end{figure}

To examine the dependence of the half-life limit on the position of
the 3$\sigma$ peak interval (12.7\,keV) in the spectrum, we shifted the peak
interval between 2028\,keV and 2048\,keV. It results in a variation of 
the half-life limit between $2.5\times10^{25}$\,y and
$1.2\times10^{25}$\,y at 90\% C.L. (for the data with pulse-shape
analysis). This demonstrates a rather smooth background in the
considered energy region. 
Figure \ref{both} shows the combined spectrum of the five detectors after 
53.93\,kg\,y and the  spectrum of point-like interactions, 
corrected for the detection efficiency,
after 35.5\,kg\,y. The solid lines represent the exclusion limits for
the two spectra at the 90\% C.L.
Using the matrix elements of \cite{staudt} and neglecting right-handed 
currents, we can
convert the lower half-life limit into an upper limit on the effective 
Majorana neutrino mass, which are listed in Table
\ref{n_mass}.

\begin{table}[t]
\begin{center}
\renewcommand{\arraystretch}{1.4}
\setlength\tabcolsep{8pt}
\begin{tabular}{l|c|c|c}
& T$_{1/2}^{0\nu} > $  & $\langle m \rangle < $  & C.L. [\%]\\
\hline
Full data set & 1.3$\times 10^{25}$\,y &  0.42\,eV & 90\\
              & 2.2$\times 10^{25}$\,y &  0.33\,eV & 68\\
SSE data      & 1.9$\times 10^{25}$\,y &  0.35\,eV & 90\\
              & 3.1$\times 10^{25}$\,y &  0.27\,eV & 68\\
\end{tabular} 
\end{center}
\caption{\label{n_mass}Limits on the effective Majorana neutrino mass from the
  \onbb{}-decay of $^{76}$Ge calculated with the matrix elements from \protect{\cite{staudt}}.}
\end{table}

The HEIDELBERG-MOSCOW  experiment is presently giving the most stringent upper limit on the
Majorana neutrino mass, of 0.35\,eV at 90\%~C.L. (0.27\,eV at 68\%~C.L.).
The values quoted in a previous paper \cite{prl}, with a statistical significance
of 24.2\,kg\,y  of data with pulse shape analysis, were 0.2\,eV for
the mass limit and  0.38\,eV for the
sensitivity of the experiment (both at 90\%~C.L.), after  the
recommendation of \cite{pdg98}. Thus, not unexpected, after additional
11.3\,kg\,y of statistics, the limit on the effective neutrino mass
approached the experimental sensitivity, as defined in \cite{feld}.
The mass limit varies within a factor of less than two for different matrix
element calculations (see the discussion in \cite {kk3}).

\section{Summary and Discussion}

\begin{figure}[t]
\begin{center}
\hspace*{-10mm}
\includegraphics[height=11.cm, angle=-90]{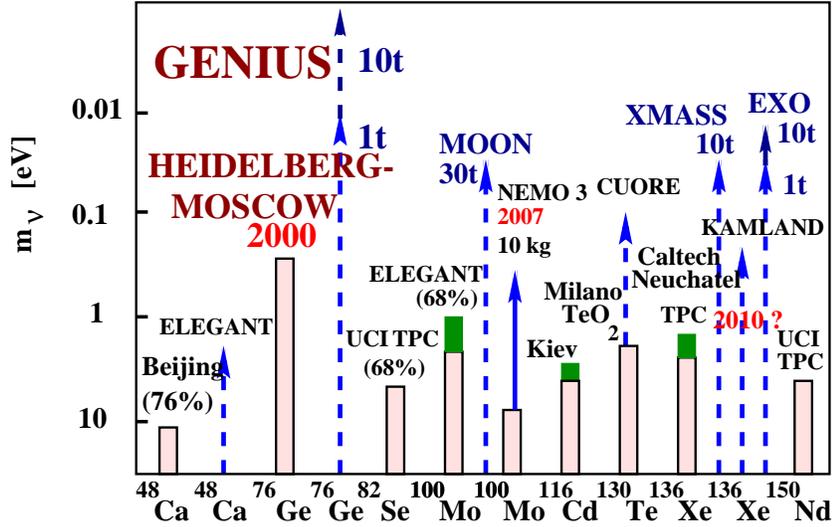}
\end{center}
\caption{Present situation, and the expectation for the future, of the 
  most promising $\beta\beta$ experiments. Light parts of the bars:
  present status; dark parts: expectation for running experiments;
  solid and dashed lines: experiments under construction and proposed
  experiments, respectively (from \cite{nanpino}).}
\label{masses}
\end{figure}

We performed an analysis of the most recent data of the
Heidelberg-Moscow double beta decay experiment. 
The data of the complete setup with  five enriched $^{76}$Ge-detectors, 
with a total statistical significance of 47.4\,kg\,y, 
were analyzed  with respect to the two-neutrino and Majoron emitting
decay modes for the first time.  A Monte Carlo simulation based on a
modified version of {\sc Geant3.21} was performed in order to identify 
the most significant background sources and to establish a
quantitative background model.
The theoretical shapes of the \tnbb{} and \mnbb{} decay spectra were
fitted in a maximum-likelihood fit to 
the resulting spectrum after subtraction of the background model from 
the measured, summed spectrum of all detectors. 
The low-energy background of the HEIDELBERG-MOSCOW experiment requires 
further investigation. A possible background source not taken into
account so far could be surface contaminations of the crystal and/or
copper parts of the cryostats with \ch{210}{Pb}, which is produced and 
accumulated by the decay of \ch{222}{Rn}. This and other potential
background sources will be implemented in a new Monte Carlo simulation 
using {\sc Geant4} \cite{geant4}. A more complete background model
will allow to determine the half-life of the $2\nu\beta\beta$ decay
with still higher precision.

\begin{figure}[t]
\begin{center}
\vspace*{15mm}
\includegraphics[height=6.cm]{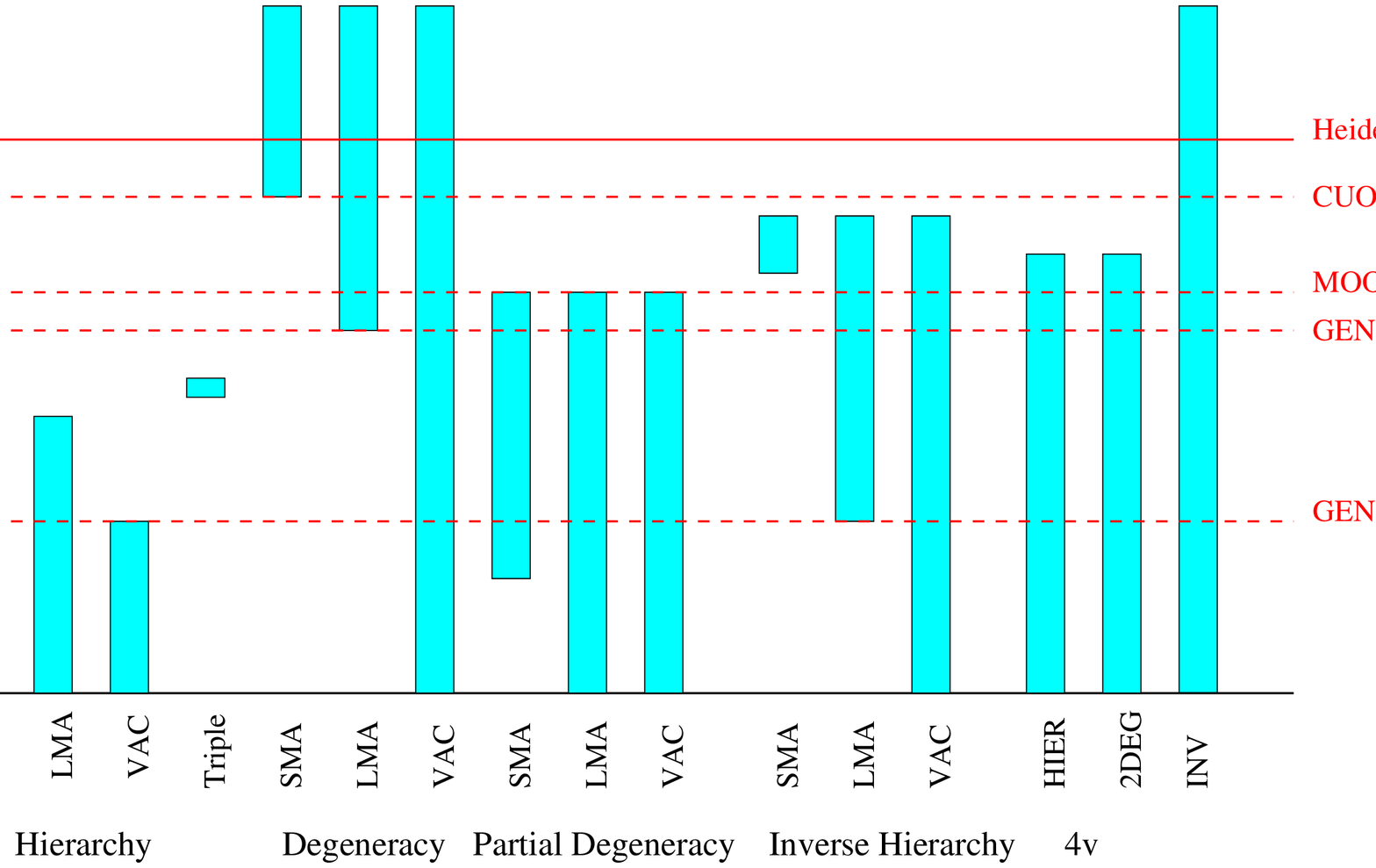}
\end{center}
\caption{Ranges of the effective Majorana neutrino-mass allowed from all
  present neutrino-oscillation experiments, for
  several neutrino-mass-scenarios, compared to our present limit (solid 
  line) and to the potential of planned future
  $\beta\beta$-experiments (dashed lines) (see \cite{paes,kk1,kk3,nanpino})}
\label{massmodels}
\end{figure}

The resulting half-life for the \tnbb{} decay confirms our previous
measurement and confirms theoretical expectations
\cite{muto,staudt} 
within a factor of two (a factor of $\sqrt{2}$ in the matrix element).
No evidence for a Majoron-accompanied decay or for the neutrinoless
decay was observed.
The upper limit on the effective Majorana neutrino mass of 0.35\,eV
(0.27\,eV) (using the matrix elements of \cite{staudt}) is the
worldwide most stringent limit up to now. 
In Fig. \ref{masses} this value from the HEIDELBERG-MOSCOW experiment
is compared with limits of the most sensitive other $\beta\beta$-experiments.
With this result for the limit of the effective Majorana neutrino mass
double beta experiments start to enter into the range to give a
serious contribution to the neutrino mass matrix (Fig. \ref{massmodels}). 

In degenerate models we can conclude from the experimental bound an
upper limit on the mass of the heaviest neutrino. For the Large Mixing 
Angle (LMA) MSW solution of the solar neutrino problem we obtain
$m_{1,2,3}<1.1$\,eV, implying $\sum_i m_i<3.2$\,eV
\cite{paes,kk1,nanpino}. This first number is sharper than what has
recently been obtained from tritium ($m<2.2$\,eV), and the second is
sharper than the limit $\sum_i m_i<5.5$\,eV still compatible with most
recent fits of Cosmic Microwave Background radiation and Large Scale
Structure data (see e.g. \cite{cmb}). 
The present sensitivity of the HEIDELBERG-MOSCOW
experiment probes cosmological models including hot dark matter already
now on a level of future satellite experiments MAP and PLANCK (see
\cite{paes,kk1}). It is of interest also for new {\em Z-burst}-models
recently discussed as explanation for super-high energy cosmic ray
events beyond the GZK cutoff \cite{wei99,paes01}.

The result for $<\!m\!>$ from the HEIDELBERG-MOSCOW experiment has found large resonance, and it has been shown that it excludes 
for example the Small Mixing Angle MSW solution of the solar neutrino problem 
in degenerate scenarios, if neutrinos are considered as hot dark matter in
the universe \cite{sma1,sma2,sma3}. This conclusion has been drawn,
before the Superkamiokande collaboration presented their evidence for
exclusion of SMA MSW solution, in June 2000.

If future searches will show, that $m_\nu>0.1$\,eV, then the three-neutrino 
mass schemes, which will survive, are those with neutrino mass
degeneracy, or four-neutrino schemes with inverse mass hierarchy (see
Fig. \ref{massmodels} and \cite{paes,kk1,nanpino}).
A substantial increase in sensitivity of double beta experiments
beyond this level, requires new experimental approaches, making use of
much higher source strength and drastically reduced background.
This could be accomplished  by our
proposed GENIUS project \cite{genius} which, operating 0.1-10 tonnes of
enriched $^{76}$Ge directly in ultrapure liquid nitrogen, could test
the effective Majorana neutrino mass down to 0.01 or even 0.002 eV.


%

\end{document}